\documentclass[12pt]{iopart}

\usepackage{graphicx}

\begin{document}

\title{Interaction between light and matter: A photon wave function approach}

\author{Pablo L Saldanha$^{1,2}$ and C H Monken$^1$}
\address{$^1$ Departamento de F\'isica, Universidade Federal de Minas
Gerais, Caixa Postal 702, 30161-970, Belo Horizonte, MG,
Brazil}%
\address{$^2$ Department of Physics, University of Oxford, Clarendon Laboratory, Oxford, OX1 3PU, United Kingdom} \ead{p.saldanha1@physics.ox.ac.uk}

\begin{abstract} The Bialynicki-Birula--Sipe photon wave function
formalism is extended to include the interaction between photons and
continuous non-absorptive media. When the second quantization of
this formalism is introduced, a new way of describing the
quantum interactions between light and matter emerges. As an example
of application, the quantum state of the twin photons generated by
parametric down conversion is obtained in agreement with previous
treatments, but with a more intuitive
interpretation.
\end{abstract}

\maketitle

\section{Introduction}

About sixty years ago, Newton and Wigner showed that there is no
position operator for a massless particle with spin higher than
$1/2$ \cite{wigner49}. This fact lead many authors to conclude that
it is not possible to define a wave function for a photon, which has
zero mass and spin (or helicity) 1. However, Bialynicki-Birula
\cite{birula94,birula96} and, independently, Sipe \cite{sipe95},
introduced a function of the position and time coordinates that
completely describes the quantum state of a photon. In their
opinion, which is shared by us, such function may be referred to as the photon
wave function. The wave equation for this function can be derived
from the Einstein kinematics for a particle with spin 1 and zero
mass in the same way that the Dirac equation is obtained for a
massive particle with spin $1/2$ \cite{raymer05,smith07}. A strong
argument in favor of this photon wave function formulation is that
the corresponding wave equation is completely equivalent to the
Maxwell equations in vacuum.


In the present work we extend the Bialynicki-Birula--Sipe formalism to include the
interaction of photons with non-absorptive continuous media. In section \ref{sec:vac} we  briefly discuss the photon wave equation in vacuum and some properties of photon wave functions. In section \ref{sec:medium} the photon wave equation is
modified including a term associated with the induced current density
in the medium, that makes it equivalent to the macroscopic Maxwell equations for
electromagnetic waves in the presence of matter. In section \ref{sec:secquant} we apply the second
quantization procedure to the photon wave function formalism in a linear non-dispersive and non-absorptive medium, allowing for  description of quantum phenomena. In section \ref{sec:scat} we present a method for treating the scattering of photons by material media. As an example of application of the proposed formalism, in section \ref{sec:cpd} the state of the photon pairs generated by
parametric down conversion is obtained with an intuitive derivation. In section \ref{sec:conc} we present our concluding remarks and summarize our results.

\section{Photon wave equation in vacuum}\label{sec:vac}

The Bialynicki-Birula--Sipe wave equation for a photon described by the 
wave function $\mathbf{\Psi}(\mathbf{r},t)$ is
\cite{birula94,birula96,sipe95} \begin{equation}\label{eq onda vac}
    \rmi\hbar\frac{\partial \mathbf{\Psi}}{\partial t}=\hbar c \hat{\sigma} \mathbf{\nabla}\times\mathbf{\Psi},
\end{equation} where $\hbar$ is the Planck's constant divided by $2\pi$ and $\hat{\sigma}$ is an operator that acts on the
photon helicity. A spin one can be represented by a vector in the
3-dimensional complex space, so the vectorial character of
$\mathbf{\Psi}$ incorporates its internal degrees of freedom.
The photon wave function can be decomposed in the eigenvectors
of the helicity operator $\hat{\sigma}$ and also in its 
real and imaginary parts: \begin{equation}\label{psi}
\mathbf{\Psi}=\mathbf{\Psi}^++\mathbf{\Psi}^-,\; \mathrm{with}\;\;\mathbf{\Psi}^{\pm}=\sqrt{\frac{\varepsilon_0}{2}} \mathbf{E}_\pm\pm
\rmi\sqrt{\frac{1}{2\mu_0}}\mathbf{B}_\pm\;\;\mathrm{and}\;\;\;\hat{\sigma}\mathbf{\Psi}^{\pm}=
\pm\mathbf{\Psi}^{\pm}.\end{equation} If we associate the real
vectors $\mathbf{E}\equiv \mathbf{E}_++\mathbf{E}_-$ and $\mathbf{B}\equiv
\mathbf{B}_++\mathbf{B}_-$ with the electric and magnetic
fields of the photon and $\varepsilon_0$ and $\mu_0$ with the
permittivity and the permeability of free space, \eref{eq onda
vac} together with the divergence-free condition of the photon wave
function $\nabla\cdot\mathbf{\Psi}=0$ (which is essential to
arrive at the photon wave equation from Einstein kinematics
\cite{raymer05,smith07}) are completely equivalent to the set of the four
Maxwell equations in vacuum. The eigenvalues of the helicity
operator correspond to photons with left and right circular
polarizations. An arbitrary polarization state can be written as a
particular combination of these two.

There is a peculiarity in the form \eref{psi} for the photon wave
function: Its modulus squared has the unit of energy density, not of
probability density, and agrees with the classical electromagnetic
energy density. This fact is associated with the problem of
non-localization of the photon \cite{wigner49}, and it happens that
the energy density is the local useful concept, since photons have no other scalar quantities such as mass or charge, but only energy \cite{sipe95}. The normalization
condition is that the integral of $|\mathbf{\Psi}|^2$ is equal
to the expectation value of the photon energy. There
is no direct association of the modulus squared of the wave function
with the probability density of detecting the photon in the region,
unless we consider dimensions larger than the photon wavelength
\cite{mandel66}. But neither the localization of massive relativistic
particles is a clear concept. In this sense, photons are
not significantly different from massive particles, as both can have an exponential falloff of their wave functions with distance in relativistic treatments \cite{birula98}, the difference being that photons are always relativistic.

\section{Photon wave equation in material media}\label{sec:medium}

Our proposal to treat the interaction of photons with continuous media
is to include a term proportional to the current density $\mathbf{J}$
induced in the media due to the presence of the photon in 
\eref{eq onda vac}: \begin{equation}\label{eq onda mat}
\rmi\hbar\frac{\partial\mathbf{\Psi}}{\partial t}=\hbar
c\hat{\sigma}\mathbf{\nabla}\times\mathbf{\Psi}-\rmi\hbar\frac{\mathbf{J}}{\sqrt{2\varepsilon_0}}\;.
\end{equation} This approach is similar to the one adopted by Keller \cite{keller00}. However, Keller was interested in calculating the wave function of a photon emitted from an atom and the current density was a source term. Here we are considering continuous media and the current density
$\mathbf{J}$ represents a response of the medium due to the
presence of the photon, not being an independent source of photons. So the objective of the present work is very different from Keller's, although the initial step is similar. When
considering continuous media, we are implicitly assuming that the
photon wavelength is much larger than the distances between the atoms that compose the
medium, in such a way that the collective motion of the charges may be
approximated by a continuous current density.  We can write
\begin{equation}\label{j clas}
	\mathbf{J}\equiv\mathbf{J}_\mathrm{f}+\mathbf{\nabla}\times
\mathbf{M}+{\partial \mathbf{P}}/{\partial t},
\end{equation}
where $\mathbf{M}$ and
$\mathbf{P}$ are the magnetization and electric polarization of the
medium and $\mathbf{J}_\mathrm{f}$ is the free current density, all of them being 
functions of the electric and magnetic fields of the photon. For instance, in a non-dispersive dielectric medium the polarization can be written as
\begin{equation}\label{pol} P_i(\mathbf{r},t)=\sum_j\chi^{(1)}_{ij}{E}_j(\mathbf{r},t)+\sum_{j,k}\chi^{(2)}_{ijk}{E}_j(\mathbf{r},t){E}_k(\mathbf{r},t)+...,
\end{equation}
expanded in linear and non-linear responses to the photon field. Since the non-linear response terms represent situations in which there are more than one photon involved, a procedure of second quantization of the photon wave function is necessary to correctly describe them. This issue will become clear in section \ref{sec:cpd} where we treat the parametric down-conversion, which arises from the second-order non-linear response of the medium. So, if we want to deal with the wave function of a single photon, we must disregard the non-linear response of the medium. In section \ref{sec:cpd} we will be interested in the interaction of photons with non-linear media after applying the second quantization procedure to the photon wave function, so we will continue referring to the current density $\mathbf{J}$ in this section  without specification of its linear or non-linear behaviors.

Using the photon wave
function from   \eref{psi}, we can see that  \eref{eq onda mat} is equivalent
to the Maxwell equations by writing its real and
imaginary parts. Its real part with the application of the operator
$\hat{\sigma}$ and its imaginary part give:  
\begin{equation}\label{max ph 1}
    \mathbf{\nabla}\times
 \mathbf{E}= -\frac{\partial \mathbf{B}}{\partial t}\;,
\end{equation}  
\begin{equation}\label{max ph 2}
    \mathbf{\nabla}\times \frac{\mathbf{B}}{\mu_0}=\varepsilon_0\frac{\partial
\mathbf{E}}{\partial t}+\mathbf{J}\;,\;\mathrm{or}\;
 \mathbf{\nabla}\times
 {\mathbf{H}}=\mathbf{J}_\mathrm{f}+\frac{\partial\mathbf{D}}{\partial
 t}\;,
\end{equation}
with $\mathbf{E}\equiv \mathbf{E}_++\mathbf{E}_-$, $\mathbf{B}\equiv
\mathbf{B}_++\mathbf{B}_-$,
$\mathbf{H}\equiv{\mathbf{B}}/{\mu_0}-\mathbf{M}$ and
$\mathbf{D}\equiv\varepsilon_0\mathbf{E}+\mathbf{P}$.  Taking the
divergence of the above equations we obtain
\begin{eqnarray} \label{max3}
&&\frac{\partial}{\partial t}(\mathbf{\nabla}\cdot\mathbf{B})=0\;.
\\\label{max4} &&\frac{\partial}{\partial
t}(\mathbf{\nabla}\cdot\mathbf{D})=-\mathbf{\nabla} \cdot
\mathbf{J}_\mathrm{f}\;,\;\mathrm{or}\;\frac{\partial}{\partial
t}(\mathbf{\nabla}\cdot\mathbf{D}-\rho_\mathrm{f})=0\;, \end{eqnarray}
where $\rho_\mathrm{f}$ is the free charge density and we have
$\mathbf{\nabla} \cdot \mathbf{J}_\mathrm{f}=-{\partial}\rho_\mathrm{f}/{\partial
t}$.  What the above equations tell us is that if
$\mathbf{\nabla}\cdot\mathbf{B}=0$ and
$\mathbf{\nabla}\cdot\mathbf{D}=\rho_\mathrm{f}$ at any particular instant
of time, which is true before the photon interacts with the media,
these relations must be always satisfied.  With this argument we
complete the proof that   \eref{eq onda mat} is 
equivalent to the set of Maxwell equations for electromagnetic waves.
So our treatment predicts that the behavior of photons in linear
non-absorptive media, where a one-photon theory may be used, is completely analogous to the behavior of
classical electromagnetic pulses.  This statement is supported by experiments \cite{steinberg92} and is difficult to
obtain with the formalism of second quantization
\cite{glauber91,barnett92,scheel08,hillery09,philbin10}. So we believe that in the same way that the Schr\"odinger equation is used to treat the behavior of massive particles under the action of a classical potential, the formalism of this section may be used to describe the interaction of photons with material media when there is no absorption or emission of photons. In cases where there is absorption or emission of photons we must apply the second quantization procedure on the photon wave function, to be done in the next section.

From   \eref{eq onda mat} we can derive a continuity equation for
the photon wave function.  Taking the scalar product with $\mathbf{\Psi}^*$ on both sides of  \eref{eq onda mat}
 and summing the complex conjugate, after verifying that $(\hat{\sigma} \nabla
    \times\mathbf{\Psi})^*\cdot\mathbf{\Psi}= (\nabla
    \times\mathbf{\Psi}^*)\cdot(\hat{\sigma}\mathbf{\Psi})$ and using the vectorial identity ${{\nabla}\cdot\left(\mathbf{A}\times
\mathbf{B}\right)} ={ \mathbf{B}\cdot(\nabla\times
\mathbf{A})-\mathbf{A}\cdot(\nabla\times \mathbf{B})}$, we arrive at
\begin{equation}\label{continuidade mat} \frac{\partial}{\partial
t}|\mathbf{\Psi}|^2=-\mathbf{\nabla}\cdot
\mathbf{S}-\frac{1}{\sqrt{2\varepsilon_0}}(\mathbf{\Psi}+\mathbf{\Psi}^*)\cdot
\mathbf{J}, \end{equation} where
$\mathbf{S}\equiv-\rmi c\,\mathbf{\Psi}^*\times(\hat{\sigma}\mathbf{\Psi})$ is the current density of the photon wave function. Using  (\ref{psi}) and taking the time average of $\mathbf{S}$ over one period of oscillation of the photon field, we have $\langle\mathbf{S}\rangle=\langle\mathbf{E}\times\mathbf{B}\rangle/\mu_0$,
analogous to the electromagnetic energy flux.  The term
$(\mathbf{\Psi}+\mathbf{\Psi}^*)\cdot\mathbf{J}/\sqrt{2\varepsilon_0}=\mathbf{E}\cdot
\mathbf{J}$ represents the density of work done
on the charges of the medium per unit time.  The continuity equation
for the photon wave function is an energy continuity equation.  In our
formalism, we are assuming that the photon energy density has always the same
form when written in terms of the photon electric and magnetic fields, independently of the medium, and that the material energy density
is calculated by the work done on the charges of the medium, without
distinction whether the work is done on bound or free currents. This approach differs from the one adopted by Bialynicki-Birula to treat photons in material media \cite{birula96}, where the photon wave function is written as $\mathbf{F}=\mathbf{D}/\sqrt{2\varepsilon}\pm\mathrm{i}\mathbf{B}/\sqrt{2\mu}$ and there is no current term in the photon wave equation, since the bound currents are included in the photon wave function via $\mathbf{D}$, $\chi_\mathrm{e}$ and $\chi_\mathrm{m}$. The photon energy density $|\mathbf{F}|^2$ is a function of the electric permittivity $\varepsilon$ and magnetic permeability $\mu$ of the medium since we have $\mathbf{D}=\varepsilon \mathbf{E}$ and
$\mathbf{H}=\mathbf{B}/\mu$ in a linear isotropic medium. 

When a
photon enters a medium, part of its energy is transferred to the
medium.  The same occurs with its momentum.  The discussion of the
correct form for the momentum and energy densities of classical
electromagnetic waves in a medium has been the subject of a long
debate \cite{pfeifer07}.  The eventual conclusion is that there are
many compatible ways of defining the electromagnetic and material
parts of the energy and momentum densities of electromagnetic waves in material
media.  Here we are assuming a particular division in the quantum
domain, where the electromagnetic parts are
$|\mathbf{\Psi}|^2=\varepsilon_0|\mathbf{E}|^2/2+|\mathbf{B}|^2/(2\mu_0)$
for the energy density, $\mathbf{S}$ for the energy flux and
$\mathbf{S}/c^2$ for the momentum density.  This division has been successfully
tested with reasonable models in the classical domain
\cite{saldanha10-mom,saldanha11-en}. 

Our formulation gives the same results as Bialynicki-Birula's for linear media: photons obey the Maxwell equations, the difference being that we implicitly divided the total energy and momentum of the photon in the medium in their electromagnetic and material parts, associating only the electromagnetic part with the photon wave function. As a consequence, our formulation can be readily extended to non-linear media after the procedure of second quantization of the photon wave function, representing an advantage to treat more general cases, as we discuss in the next sections.

\section{Second quantization of the photon wave function in a linear non-dispersive and non-absorptive medium} \label{sec:secquant}

The second quantization
procedure can be applied to the one-photon theory to build a quantum
field theory for describing states with an arbitrary number of photons.
This procedure is described in  \cite{smith07}. Here we will make a similar second quantization procedure in an infinite linear non-dispersive and non-absorptive medium. In such a medium, the material acquires an energy density $u_{\mathrm{mat}}=\chi_{\mathrm{e}}\varepsilon_0|\mathbf{E}|^2/4+\chi_{\mathrm{m}}|\mathbf{B}|^2/[4(1+\chi_{\mathrm{m}})\mu_0]$ due to the presence of the photon  \cite{saldanha11-en}, $\chi_\mathrm{e}$ and $\chi_\mathrm{m}$ being the electric and magnetic susceptibilities of the medium, apart from the electromagnetic energy density $|\mathbf{\Psi}|^2=\varepsilon_0|\mathbf{E}|^2/2+|\mathbf{B}|^2/(2\mu_0)$. So the total energy density of the photon in such a medium can be written as
$|\mathbf{\Psi}'|^2$, where $\mathbf{\Psi}'=\mathbf{\Psi'}^{+}+\mathbf{\Psi'}^{-}$ is the ``dressed photon'' wave function, with \begin{equation}\label{psil}
    \mathbf{\Psi'}^{\pm}=\sqrt{\frac{\varepsilon}{2}}
\mathbf{E}_\pm\pm
\rmi\sqrt{\frac{1}{2\mu}}\mathbf{B}_\pm\;\;\mathrm{and}\;\;\hat{\sigma}\mathbf{\Psi'}^{\pm}=
\pm\mathbf{\Psi'}^{\pm}, \end{equation} where
$\varepsilon\equiv(1+\chi_\mathrm{e})\varepsilon_0$ and $\mu\equiv(1+\chi_\mathrm{m})\mu_0$. It is straightforward to show that $|\mathbf{\Psi'}|^2=|\mathbf{\Psi}|^2+u_{\mathrm{mat}}$. This form for $\mathbf{\Psi'}$ is equivalent to the one used by Bialynicki-Birula in the case of non-dispersive linear media \cite{birula96}. However, if dispersion is included the modulus squared of the Bialynicki-Birula's form for the photon wave function in the medium does not give the total energy density anymore, while the division of the total energy density in electromagnetic and material parts remains valid \cite{saldanha11-en}. Although we will not treat dispersive media here, it is important to stress this fundamental difference between the treatments. The above relations can be inverted, giving
\begin{eqnarray}\label{e-psil}
    \mathbf{E}&\equiv&\mathbf{E}_++\mathbf{E}_-=\frac{1}{\sqrt{2\varepsilon}}(\mathbf{\Psi}'+\mathbf{\Psi'}^{*})\;\;,\\ \label{b-psil}
    \mathbf{B}&\equiv&\mathbf{B}_++\mathbf{B}_-=-\rmi{\sqrt{\frac{\mu}{2}}}[(\hat{\sigma}\mathbf{\Psi}')-(\hat{\sigma}\mathbf{\Psi}')^*]\;.
\end{eqnarray}
The total energy of the ``dressed photon'' can be written as \begin{equation}\label{en foton
linear}
    H=\int \rmd^3r|\mathbf{\Psi}'(\mathbf{r},t)|^2\;.
\end{equation}
The ``dressed photon'' wave function can be decomposed in its energy eigenstates, which are plane-wave modes:
\begin{equation}\label{psil r-p}
    {\mathbf{\Psi}}'(\mathbf{r},t)=\sum_s\int{\rmd^3k}\sqrt{\frac{\hbar\omega}{(2\pi)^3}}\;a_{\mathbf{k}s}\;\mathrm{e}^{\rmi(\mathbf{k}\cdot
    \mathbf{r}-\omega t)}\;\mathbf{\hat{e}}_{\mathbf{k}s}
\end{equation}
with $\sum_s\int \rmd^3k\;|a_{\mathbf{k}s}|^2=1$, where the plane-wave modes are defined by their wavevector $\mathbf{k}$ and polarization vector $\mathbf{\hat{e}}_{\mathbf{k}s}$, with frequency given by $\omega=ck/n$,  $n=\sqrt{(1+\chi_\mathrm{e})(1+\chi_\mathrm{m})}$ being the refraction index of the medium. Substituting  (\ref{psil r-p}) in  (\ref{en foton
linear}), the energy can be written as
\begin{equation}\label{en foton linear autoestados}
    H=\sum_s\int \rmd^3k\; \hbar\omega|a_{\mathbf{k}s}|^2\;.
\end{equation}

The above Hamiltonian is equivalent to the Hamiltonian of a continuous set of harmonic oscillators \cite{mandel}, and the result of the second quantization procedure is to raise the ``dressed photon'' wave
function \eref{psil r-p} to the status of a field operator
\begin{equation}\label{op psil r-p}
    \hat{\mathbf{\Psi}}'(\mathbf{r},t)=\sum_s\int{\rmd^3k}\sqrt{\frac{\hbar\omega}{(2\pi)^3}}\;\mathrm{e}^{\rmi(\mathbf{k}\cdot
    \mathbf{r}-\omega t)}\;\hat{a}_{\mathbf{k}s}\mathbf{\hat{e}}_{\mathbf{k}s}\;,
\end{equation}
where $\hat{a}_{\mathbf{k}s}$ is the annihilation operator for photons in the mode $(\mathbf{k},s)$ that, together with the creation operators $\hat{a}^\dag_{\mathbf{k}s}$, obey the commutation relations
\begin{equation}\label{rel comut}\nonumber
    [\hat{a}_{\mathbf{k}s},\hat{a}_{\mathbf{k}'s'}]=[\hat{a}_{\mathbf{k}s}^\dag,\hat{a}^\dag_{\mathbf{k}'s'}]=0,\;\;
    [\hat{a}_{\mathbf{k}s},\hat{a}^\dag_{\mathbf{k}'s'}]=\delta^3(\mathbf{k}-\mathbf{k}')\delta_{ss'}.
\end{equation}
According to (\ref{e-psil}) and (\ref{b-psil}), the electric and magnetic fields also become operators\footnote{To arrive at  (\ref{op b}) from  (\ref{b-psil}), we must use the fact that $\hat{\sigma}\mathbf{\hat{e}}_{\mathbf{k}\sigma}=i(\mathbf{k}/k)\times\mathbf{\hat{e}}_{\mathbf{k}\sigma}$, where $\mathbf{\hat{e}}_{\mathbf{k}\sigma}$ is a polarization vector with helicity $\sigma$. This can be verified choosing $\mathbf{k}/k=\mathbf{\hat{z}}$,
$\mathbf{\hat{e}}_{\mathbf{k}+}=(\mathbf{\hat{x}}+i\mathbf{\hat{y}})/\sqrt{2}$
and
$\mathbf{\hat{e}}_{\mathbf{k}-}=(\mathbf{\hat{x}}-i\mathbf{\hat{y}})/\sqrt{2}$. }: 
\begin{equation}\label{op e}
    \hat{\mathbf{E}}(\mathbf{r},t)=
    \sum_s\int{\rmd^3k}\sqrt{\frac{\hbar\omega}{2\varepsilon(2\pi)^3}}\;
    \mathrm{e}^{\rmi(\mathbf{k}\cdot \mathbf{r}-\omega
    t)}\,\hat{a}_{\mathbf{k}s}\,\mathbf{\hat{e}}_{\mathbf{k}s}\;+\;\mathrm{h.c.},
\end{equation}
\begin{equation}\label{op b}
    \hat{\mathbf{B}}(\mathbf{r},t)=
    \sum_s\int{\rmd^3k}\sqrt{\frac{\hbar\omega\mu}{2(2\pi)^3}}\;
    \mathrm{e}^{\rmi(\mathbf{k}\cdot \mathbf{r}-\omega
    t)}\,\hat{a}_{\mathbf{k}s}\left[\frac{\mathbf{k}}{k}\times\mathbf{\hat{e}}_{\mathbf{k}s}\right]+\;\mathrm{h.c.}.
\end{equation}

Equations (\ref{op e}) and (\ref{op b}) for the electric and magnetic field operators are identical to the ones obtained with the traditional second quantization of the electromagnetic field in vacuum if we substitute $\varepsilon$ by $\varepsilon_0$ and  $\mu$ by $\mu_0$ \cite{mandel}. In the treatment of Smith and Raymer for the quantization of the photon wave function in vacuum \cite{smith07}, 
the operator $\hat{\mathbf{\Psi}}$ they find is equivalent to our $\hat{\mathbf{\Psi}}'+\hat{\mathbf{\Psi}}'^\dagger$, where  $\hat{\mathbf{\Psi}}'$ is given by (\ref{op psil r-p})  with $\chi_\mathrm{e}=\chi_\mathrm{m}=0$. We believe that our form, in which $\hat{\mathbf{\Psi}}'$ has only annihilation operators, is more convenient, since it agrees with the condition \begin{equation}\label{op psil-e o}
    \hat{\mathbf{\Psi}}'=\sqrt{\frac{\varepsilon}{2}}\hat{\mathbf{E}}+\rmi\hat{\sigma}\sqrt{\frac{1}{2\mu}}\hat{\mathbf{B}}
\end{equation} derived from (\ref{psil}).

A one-photon state can be written as 
\begin{equation}\label{exemplo estado 1}
    |\Psi'^{(1)}\rangle=\sum_s\int {\rmd^3k}
    \tilde{\Psi}'^{(1)}_s(\mathbf{k})\hat{a}^\dag_{\mathbf{k}s}|\mathrm{vac}\rangle\;,
\end{equation} with
$\sum_s\int {\rmd^3k}
    |\tilde{\Psi}'^{(1)}_s(\mathbf{k})|^2=1$ and $|\mathrm{vac}\rangle$ representing the vacuum state of the electromagnetic field. The corresponding
one-photon wave  function is \cite{smith07}
\begin{equation}\label{psi 1
foton}
    \mathbf{\Psi'}^{(1)}(\mathbf{r},t)\equiv
\langle \mathrm{vac}
|\hat{\mathbf{\Psi}}\mathbf{'}(\mathbf{r},t)|\Psi'^{(1)}\rangle
=\sum_s\int {\rmd^3k}\sqrt{\frac{\hbar\omega}{(2\pi)^3}}\; \tilde{\Psi}'^{(1)}_s(\mathbf{k})
     \mathrm{e}^{\rmi(\mathbf{k}\cdot \mathbf{r}-\omega
     t)}\mathbf{\hat{e}}_{\mathbf{k}s},
\end{equation}
in agreement with (\ref{psil r-p}). 

A
two-photon state can be written as  
\begin{equation}\label{exemplo estado 2}
    |\Psi'^{(2)}\rangle=\sum_{s,s'}\int {\rmd^3k}\int {\rmd^3k'}\;    \tilde{\Psi}'^{(2)}_{s,s'}(\mathbf{k},\mathbf{k}')\hat{a}^\dag_{\mathbf{k}s}\hat{a}^\dag_{\mathbf{k}'s'}|\mathrm{vac}\rangle
\end{equation}
and the corresponding two-photon wave function is \cite{smith07,smith06}
\begin{eqnarray}\label{funcao onda 2 fotons}
\nonumber \mathbf{\Psi'}^{(2)}(\mathbf{r}_1,\mathbf{r}_2,t) &\equiv&
\langle \mathrm{vac}
|\hat{\mathbf{\Psi}}\mathbf{'}(\mathbf{r}_1,t)\hat{\mathbf{\Psi}}\mathbf{'}(\mathbf{r}_2,t)|\Psi^{(2)}\rangle\\
&=&\sum_{s,s'}\int \frac{\rmd^3k}{{(2\pi)^{3/2}}}\int \frac{\rmd^3k'}{(2\pi)^{3/2}}
   \nonumber \left[\tilde{\Psi}'^{(2)}_{s,s'}(\mathbf{k},\mathbf{k}')+\tilde{\Psi}'^{(2)}_{s',s}(\mathbf{k}',\mathbf{k})\right]\\
    &&\times\sqrt{\hbar\omega}\mathrm{e}^{\rmi(\mathbf{k}\cdot \mathbf{r}_1-\omega t)}\mathbf{\hat{e}}_{\mathbf{k}s}\otimes
    \sqrt{\hbar\omega'}\mathrm{e}^{\rmi(\mathbf{k}'\cdot \mathbf{r}_2-\omega'
    t)}\mathbf{\hat{e}}_{\mathbf{k}'s'}.
\end{eqnarray}
The tensor product in the above equation indicates the composition of two one-photon space states. Note that the wave function is automatically symmetrized via the term $\left[\tilde{\Psi}'^{(2)}_{s,s'}(\mathbf{k},\mathbf{k}')+\tilde{\Psi}'^{(2)}_{s',s}(\mathbf{k}',\mathbf{k})\right]$. The extension to higher dimensions is straightforward.

\section{Scattering of photons}\label{sec:scat}

Here we want to find the behavior of the operators $\hat{\mathbf{E}}$ and $\hat{\mathbf{B}}$ from  (\ref{op e}) and
(\ref{op b}) when, besides the linear medium with electric and magnetic susceptibilities $\chi_\mathrm{e}$ and $\chi_\mathrm{m}$ occupying all space, there is another medium in a finite region with different properties. Our treatment can be seen as an approximation of Keller's treatment for the photon scattering \cite{keller00,keller01} in the far-field regime. Applying the second quantization procedure including the interaction with matter, the operators   $\hat{\mathbf{\Psi}}$ will obey the same wave equation
(\ref{eq onda mat}) as the wave functions ${\mathbf{\Psi}}$. So the operators $\hat{\mathbf{E}}$ and $\hat{\mathbf{B}}$ will also obey  (\ref{max ph 1}) and (\ref{max ph 2}). Equation (\ref{max ph 2}) can be written in the following way
\begin{equation}\label{max ph 2 b}
    \mathbf{\nabla}\times \frac{\hat{\mathbf{B}}}{\mu}=\varepsilon\frac{\partial
\hat{\mathbf{E}}}{\partial t}+\hat{\mathbf{J}}'\;, \end{equation}
 where
\begin{equation}
{\mathbf{J}}'\equiv{\mathbf{J}}-\nabla\times\left[\frac{\chi_\mathrm{m}{\mathbf{B}}}{(1+\chi_\mathrm{m})\mu_o}\right]-\frac{\partial(\chi_\mathrm{e}{\mathbf{E}})}{\partial t}
\end{equation}
contain terms due to the difference of the behavior of the scattering medium in relation to the quantization medium, according to (\ref{j clas}). Because the current density $\mathbf{J}'$ is a function of the electric and magnetic fields, when the fields become operators $\hat{\mathbf{J}}'$ also becomes an operator.
 
Differentiating  (\ref{max ph 2 b}) with respect to time,
using (\ref{max ph 1}) and the vectorial identity
$\nabla\times(\nabla\times \mathbf{E})=\nabla(\nabla\cdot
\mathbf{E})-\nabla^2\mathbf{E}$, we obtain a wave equation for the operator $\hat{\mathbf{E}}$:
\begin{equation}\label{eq onda
source E}
    \nabla^2\hat{\mathbf{E}}-\frac{n^2}{c^2}\frac{\partial^2\hat{\mathbf{E}}}{\partial
    t^2}=-4\pi\hat{\mathbf{f}}^\mathrm{e}\;,\;\mathrm{with}\;\hat{\mathbf{f}}^\mathrm{e}=-\mu_0
    \frac{\partial \hat{\mathbf{J}}'}{\partial
    t}-\nabla(\nabla\cdot\hat{\mathbf{E}}).
\end{equation}  The above equation is a wave equation in which 
$\hat{\mathbf{f}}^\mathrm{e}$ is the source. An analog equation can be found for the operator $\hat{\mathbf{B}}$ in a similar way: \begin{equation}\label{eq onda source B}
    \nabla^2\hat{\mathbf{B}}-\frac{n^2}{c^2}\frac{\partial^2\hat{\mathbf{B}}}{\partial    t^2}=-4\pi\hat{\mathbf{f}}^\mathrm{b}\;,\;\mathrm{with}\;\hat{\mathbf{f}}^\mathrm{b}=\frac{1}{\mu_0}\nabla\times\hat{\mathbf{J}}'.
\end{equation}

Solutions of (\ref{eq onda source E}) can be obtained using Green functions \cite{jackson}. If we want to study the scattering of a photon in a known state in the paraxial regime, where the dependence of the polarization with the propagation direction can be disregarded and a scalar theory may be used, the relevant solutions of (\ref{eq onda source
E}) for the operator $\hat{\mathbf{E}}(\mathbf{r},t)$ are
 \begin{equation}\label{psi out}
    \hat{\mathbf{E}}_{\mathrm{out}}(\mathbf{r},t)=\hat{\mathbf{E}}_{\mathrm{in}}(\mathbf{r},t)+
    \int \rmd^3r'\int_{-\infty}^t \rmd t'G^{(+)}(\mathbf{r},t;\mathbf{r}',t')\hat{\mathbf{f}}^\mathrm{e}(\mathbf{r}',t'),
\end{equation} \begin{equation}\label{G}
   \mathrm{with}\;\; G^{(\pm)}(\mathbf{r},t;\mathbf{r}',t')=\frac{c}{8\pi^2}\int_{-\infty}^{+\infty}\rmd k\frac{\mathrm{e}^{\pm \rmi k|\mathbf{r}-\mathbf{r}'|}}{|\mathbf{r}-\mathbf{r}'|}\mathrm{e}^{-\rmi\omega
    (t-t')},
\end{equation} where $\omega=kc/n$. 


Integrating  (\ref{G}) in $k$, we see that the Green functions can be written as 
\begin{equation}\label{G delta}
    G^{(\pm)}(\mathbf{r},t;\mathbf{r}',t')=\frac{c}{4\pi|\mathbf{r}-\mathbf{r}'|}\;\delta(|\mathbf{r}-\mathbf{r}'|\mp
    (t-t')c/n).
\end{equation} It can also be shown, by a direct integration on the variables $\phi$ and $\theta$ in spherical coordinates $(\phi,\theta,k)$ \cite{mandel}, that  \begin{equation}\label{G2'}\nonumber
    \frac{2\rmi c}{(2\pi)^3}\int \frac{\rmd^3k}{k}\mathrm{e}^{\rmi[\mathbf{k}\cdot
    (\mathbf{r}-\mathbf{r}')-\omega (t-t')]}=G^{(+)}(\mathbf{r},t;\mathbf{r}',t')-G^{(-)}(\mathbf{r},t;\mathbf{r}',t').
\end{equation} If we are studying the scattering of a photon, the observation time $t$ will always be greater than the time of the interaction $t'$. So, according to
(\ref{G delta}), $G^{(-)}(\mathbf{r},t;\mathbf{r}',t')=0$ and we can write \begin{equation}\label{G2}
    G^{(+)}(\mathbf{r},t;\mathbf{r}',t')=\frac{2\rmi c}{(2\pi)^3}\int \frac{\rmd^3k}{k}\mathrm{e}^{\rmi[\mathbf{k}\cdot
    (\mathbf{r}-\mathbf{r}')-\omega (t-t')]}.
\end{equation}

The utilization of the Green function method to predict the behavior of the electric field operator, according to  (\ref{psi out}), is very useful for calculating the state of scattered photons by a localized material medium. In linear media, we can use the same equation substituting the operator by the electric field of the photons, obtaining equivalent results to the scattering of an electromagnetic wave, since (\ref{psi out}) is obtained from the Maxwell equations. But the interesting results come from the treatment of problems that cannot be obtained from classical electrodynamics. In the next section, we will apply the treatment to the scattering of a photon by a non-linear medium that can generate a photon pair in an entangled state. We will show that the quantum state of the generated photon pair we obtain is the same as the one previously obtained \cite{hong85,mandel,monken98}, but with a more intuitive interpretation.

\section{Treatment of parametric down conversion}\label{sec:cpd}

The parametric down conversion corresponds to the conversion of one incident photon in two by a non-linear birefringent crystal. As an example of application of the formalism developed in the previous section, we will calculate the quantum state of the photon pair generated in the process of type-I parametric down conversion, in which both generated photons have polarization orthogonal to the incident photon. To simplify the treatment, we will consider that the non-linear crystal is immersed in a medium with the same linear responses as the crystal. So the medium where the field is quantized occupies all space, and the non-linearities of the crystal will be responsible for the scattering processes. The magnetic responses of these crystals are very small, so we can consider a refraction index $n=\sqrt{(1+\chi_\mathrm{e})}$. We will also disregard the dependence of the refraction index with the propagation direction and the effects of dispersion. These approximations are consistent with the observation of photons at small angles in relation to the propagation direction of the incident photon and to the use of interference filters on the detectors, post-selecting states with relatively narrow frequency spectra.

With all these considerations, the non-linear components of the polarization of the crystal can be written as
\begin{equation}\label{pol nao
linear}
    \hat{P}_i^{NL}(\mathbf{r},t)=\sum_{j,k}\chi^{(2)}_{ijk}\hat{E}_j(\mathbf{r},t)\hat{E}_k(\mathbf{r},t)\;,
\end{equation}
where terms beyond the second order were disregarded. In type-I parametric down conversion, the generated photons have ordinary polarization and the incident photon has extraordinary polarization. According to (\ref{j clas}), the non-linear current density associated with the ordinary polarization can be written as $\hat{{J}}^{'NL}_{o}=\partial\hat{P}_{o}^{NL}/\partial t$, and the scattering term, according to (\ref{eq onda source E}), is $\hat{{f}}^{\mathrm{e}}_{o}=-\mu_0\partial{J}^{'NL}_{o}/\partial t$. So the relevant source term for the scattering in (\ref{psi
out}) is
\begin{equation}\label{f cpd}
\hat{{f}}^{\mathrm{e}}_{o}(\mathbf{r}',t')=-\mu_0\frac{\partial^2\hat{P}_{o}^{NL}}{\partial
t'^2}=-\mu_0\frac{\partial^2}{\partial
t'^2}\left[\chi'^{(2)}_{ooe}(\mathbf{r}')\hat{E}_{o}(\mathbf{r}',t')\hat{E}_{e}(\mathbf{r}',t')\right],
\end{equation} where the subscript $e$ ($o$) denotes extraordinary (ordinary) polarization and we have
$\chi'^{(2)}_{ooe}\equiv\chi^{(2)}_{ooe}+\chi^{(2)}_{oeo}$, since
$\hat{E}_o(\mathbf{r}',t')$ and $\hat{E}_e(\mathbf{r}',t')$ are operators that act on modes with orthogonal polarization and therefore commute. According to (\ref{funcao onda 2 fotons}), if we denote the state of the incident photon by $|\psi^{(i)}\rangle$, the two-photon component with ordinary polarization of the output state is
\begin{equation}
{\Psi'}^{(2)}_{oo}(\mathbf{r}_1,\mathbf{r}_2,t)= \langle
\mathrm{vac} |\hat{{\Psi}}_{\mathrm{out}\,
o}'(\mathbf{r}_1,t)\hat{{\Psi}}_{\mathrm{out}\,o}'(\mathbf{r}_2,t)|\psi^{(i)}\rangle\;.
\end{equation} Considering a first order scattering, in which one of the operators $\hat{{\Psi}}_{\mathrm{out}\,o}'(\mathbf{r},t)$ above is substituted via (\ref{op psil-e o}) by the scattered electric field of (\ref{psi out}) and the other is kept with the non-scattered field, we can write
\begin{equation}\label{psi 2}
{\Psi'}^{(2)}_{oo}(\mathbf{r}_1,\mathbf{r}_2,t)= \langle
\mathrm{vac}
|\hat{{\Psi}}_{o}'(\mathbf{r}_1,t)n\sqrt{\frac{\varepsilon_0}{2}}\hat{{E}}_{\mathrm{out}\,o}(\mathbf{r}_2,t)|\psi^{(i)}\rangle
\end{equation} plus a term with the permutation 
$\mathbf{r}_1\leftrightarrow\mathbf{r}_2$, that correctly symmetrizes the wave function. From now on we will omit this permutation term to simplify the notation. The refraction index $n$ appears in the above equation because we have $\sqrt{\varepsilon}=\sqrt{(1+\chi_{\mathrm{e}})\varepsilon_0}=n\sqrt{\varepsilon_0}$. $\hat{{E}}_{\mathrm{out}\,o}$ is given by (\ref{psi out}), with $\hat{\mathbf{f}}^\mathrm{e}$ given by (\ref{f cpd}).

The state $|\psi^{(i)}\rangle$ has one photon with extraordinary polarization and in (\ref{psi 2}) 2 operators that act on ordinary polarization modes are applied to the state $|\psi^{(i)}\rangle$. We can substitute these operators by their commutator plus terms that result in zero when applied on $|\psi^{(i)}\rangle$. Using (\ref{op psil r-p}),  (\ref{rel comut}) and (\ref{op
e}), the relevant commutator to the problem in the limit that we disregard the dependence of the polarization with the propagation direction is \cite{mandel}
\begin{eqnarray}\label{comut E}\nonumber
  && \left[\hat{{\Psi}}'_o(\mathbf{r}_1,t_1),\sqrt{\frac{\varepsilon_0}{2}}\hat{E}_{o}(\mathbf{r}',t')\right]=
    \int {\rmd^3k}\frac{\hbar\omega}{2n(2\pi)^3}\mathrm{e}^{\rmi[\mathbf{k}\cdot
    (\mathbf{r}-\mathbf{r}')-\omega (t-t')]}=\\  &&=\frac{-\rmi\hbar}{4c^2}
    \frac{\partial^2}{\partial t \partial
    t'}G^{(+)}(\mathbf{r}_1,t;\mathbf{r}',t'),
\end{eqnarray}
where 
(\ref{G2}) was used in the last equality.

Inserting the value of (\ref{f cpd}) for $\hat{\mathbf{f}}^\mathrm{e}$ in
$\hat{{E}}_{\mathrm{out}\,o}(\mathbf{r}_2,t)$, obtained from (\ref{psi out}), in (\ref{psi 2}) and using 
(\ref{comut E}) to substitute the product of the operators
$\hat{{\Psi}}'_o(\mathbf{r}_1,t_1)$ and
$\sqrt{\varepsilon_0/2}\hat{E}_{o}(\mathbf{r}',t')$ by their commutator plus terms that result in zero when applied on
$|\psi^{(i)}\rangle$, we have 
\begin{eqnarray}\nonumber
     {\Psi}'^{(2)}_{oo}(\mathbf{r}_1,\mathbf{r}_2,t) =&& \frac{\rmi \hbar \mu_0}{4c^2} \int \rmd^3r'\int_{-\infty}^t \rmd t'\chi'^{(2)}_{ooe}(\mathbf{r}')G^{(+)}(\mathbf{r}_2,t;\mathbf{r}',t')\times\\&&\times\frac{\partial^2}{\partial t'^2}\left\{\left[\frac{\partial^2}{\partial t \partial
     t'}G^{(+)}(\mathbf{r}_1,t;\mathbf{r}',t')\right]E_3'(\mathbf{r}',t')\right\}\;,
   \end{eqnarray}   
with $E_3'(\mathbf{r}',t')\equiv \langle \mathrm{vac}
|n\hat{E}_{e}(\mathbf{r}',t')|\psi^{(i)}\rangle$. Substituting (\ref{G2}) and defining $E_3'(\mathbf{r}',t')\equiv\int
d^3k_3 \tilde{E}_3(\mathbf{k}_3)\exp(\rmi\mathbf{k}_3\cdot \mathbf{r}'
-i\omega_3 t)$, we have \begin{eqnarray}\label{estado cpd}\nonumber
  &&{\Psi'}^{(2)}_{oo}(\mathbf{r}_1,\mathbf{r}_2,t) = \frac{i\hbar\mu_0}{(2\pi)^6}\int \rmd^3r'\int_{-\infty}^t \rmd t' \int \rmd^3k_1\int \rmd^3k_2\int \rmd^3k_3
  \frac{\omega_1^2}{k_1} \frac{(\omega_3-\omega_1)^2}{k_2}\times\\
  &&\times \chi'^{(2)}_{ooe}(\mathbf{r}')\tilde{E}_3(\mathbf{k}_3)
\mathrm{e}^{\rmi(\mathbf{k}_3-\mathbf{k}_1-\mathbf{k}_2)\cdot\mathbf{r}'-\rmi(\omega_3-\omega_1-\omega_2)t' }
\mathrm{e}^{\rmi[\mathbf{k}_1\cdot \mathbf{r}_1-\omega_1
t]}\mathrm{e}^{\rmi[\mathbf{k}_2\cdot \mathbf{r}_1-\omega_2 t]}\;.
\end{eqnarray} 
If we observe the generated photons at a distance from the crystal such that the time light takes to go from the crystal to the detectors is greater than the duration of the incident pulse, the integral in $t'$ may be extended until $t'=\infty$, resulting in a term proportional to $\delta (\omega_3-\omega_1-\omega_2)$. So the term
$(\omega_3-\omega_1)^2$ in the above equation may be substituted by
$\omega_2^2$. The above form for the wave function of the twin photons generated by parametric down conversion is equivalent to previous treatments that consider a Hamiltonian evolution of the quantum state of light interacting with a non-linear medium using perturbation theory \cite{hong85,mandel}.

If we use the form from  (\ref{G}) to the Green function, we arrive at a more intuitive form for the quantum state of the twin photons: 
\begin{eqnarray}\label{psi final}\nonumber
    \Psi'^{(2)}_{oo}(\mathbf{r}_1,\mathbf{r}_2,t)=&&-\frac{\rmi\hbar\mu_0}{4(8\pi^2)^2}\int \rmd^3r'\int_{-\infty}^t
    \rmd t'\int \rmd k_1 \int \rmd k_2 {\chi'}_{ooe}^{(2)}(\mathbf{r}')
    {E}_{3}'(\mathbf{r}',t')\times\\&&\times \omega_1^2\frac{\mathrm{e}^{\rmi[k_1|\mathbf{r}_1-\mathbf{r}'|-\omega_1(t-t')]}}{|\mathbf{r}_1-\mathbf{r}'|}\omega_2^2\frac{\mathrm{e}^{\rmi[k_2|\mathbf{r}_2-\mathbf{r}'|-\omega_2(t-t')]}}{|\mathbf{r}_2-\mathbf{r}'|}
    \;.
\end{eqnarray}

We can interpret the above wave function as being the result of the coherent scattering of one photon in two by all non-linear scatterers of the crystal. In each one of these processes, the incident photon generates a pair of photons in spherical wave modes emitted from each non-linear scatterer. The energy and momentum conservation in the process may be seen as consequences of the fundamental indistinguishability of all possible instants of time  and  positions in which a photon pair could be generated, which forces us to coherently add all the probability amplitudes. These probability amplitudes interfere destructively when energy is not conserved. However, although the sum of the energies of the twin photons is the energy of the incident photon, the energy of each photon can assume a wide range of values. So, the photon pair is highly entangled in energy. The probability amplitudes also interfere making the sum of the momenta of the twin photons be associated with the momentum of the incident photon, depending on the crystal geometry since the crystal may absorb part of the momentum in the process. The momentum of each photon can also assume a wide range of values, so the generated photons are also entangled in linear momentum.

Comparing  (\ref{estado cpd}) and (\ref{funcao onda 2
fotons}), we see that the post-selected wave function of the twin photons on the wavevector space is proportional to
\begin{eqnarray}\nonumber\label{2photon wavenumber}
    \tilde{\Psi}^{(2)}_{oo}(\mathbf{k}_1,\mathbf{k}_2)\propto&&\frac{\sqrt{\omega_1\omega_2}}{n_1n_2}\int
    \rmd t'\int \rmd^3r' \int \rmd^3k_3\,\chi'^{(2)}_{ooe}(\mathbf{r}')\times\\&&\times\tilde{E}_3(\mathbf{k}_3)\mathrm{e}^{\rmi(\mathbf{k}_3-\mathbf{k}_1-\mathbf{k}_2)\cdot\mathbf{r}'-\rmi(\omega_3-\omega_1-\omega_2)t'
    }.
\end{eqnarray} Let us consider that the incident photon is in a beam mode that propagates in the $\mathbf{\hat{z}}$ direction and that the crystal is 
a rectangular box with dimensions much larger than the beam waist in the $x$ and $y$ directions and a small width in the $z$ direction. Let us also treat the collinear regime, in which the crystal is oriented to maximize the production of photons with wavevectors near the $\mathbf{\hat{z}}$ direction.  So the integrals in $x$ and $y$ in  (\ref{2photon wavenumber}) may be extended to infinity and the integral in $z$ results in a constant. To illustrate the energy and momentum entanglement of the state, let us represent the vectors $\mathbf{k}$
as $\mathbf{k}=\mathbf{q}+k_z\mathbf{\hat{z}}$ and
$\tilde{E}_3(\mathbf{k}_3)$ as
$\tilde{E}_3(\omega_3,\mathbf{q}_3)$, since $\omega_3$ and
$\mathbf{q}_3$ determine $\mathbf{k}_3$. Let us also substitute the integral in the $\mathbf{\hat{z}}$ component of $\mathbf{k}_{3}$
by an integral in  $\omega_3$. With all these approximations, the integral in  $t'$ gives $\delta(\omega_3-\omega_1-\omega_2)$
and the integrals on $x'$ and $y'$ give
$\delta^2(\mathbf{q}_3-\mathbf{q}_1-\mathbf{q}_2)$. So the 2-photon wave function on wavevector space can be written within the paraxial domain ($|\mathbf{q}|\ll|\mathbf{k}|$) as 
\cite{monken98} \begin{equation}
    \tilde{\Psi}^{(2)}_{oo}(\omega_1,\mathbf{q}_1,\omega_2,\mathbf{q}_2)\propto{\sqrt{\omega_1\omega_2}}\;
    \tilde{E}_3(\omega_1+\omega_2,\mathbf{q}_1+\mathbf{q}_2).
\end{equation} We cannot write this wave function as a product of a wave function for photon 1 and a wave function for photon 2, what characterizes the entanglement of the state in both energy and momentum.

\section{Conclusions}\label{sec:conc}

We extended the Bialynicki-Birula--Sipe photon wave function formalism to include the interaction between photons and continuous non-absorptive media. The photon wave equation becomes equivalent to the macroscopic Maxwell equations in material media. Applying the second quantization procedure to the photon wave function, we obtained an alternative treatment for the quantum interactions between light and continuous media, that can be specially useful to treat the scattering of light by matter. As an example of application of the proposed formalism, the quantum state of the twin photons generated by parametric down conversion was obtained in agreement with previous treatments, but with a more intuitive interpretation.

\ack
The authors acknowledge Brian J. Smith for useful discussions. This work was supported by the Brazilian agencies CNPq and CAPES.


\end{document}